\magnification=\magstep1

% for phys rev lett
%
\hsize = 6.50truein
\vsize = 8.50truein
\hoffset = 0.0truein
\voffset = 0.0truein
\lineskip = 2pt
\lineskiplimit = 2pt
\overfullrule = 0pt
\tolerance = 2000
\topskip = 0pt
\baselineskip = 18pt
\parindent = 0.4truein
\parskip = 0pt plus1pt
\def\medskip{\vskip6pt plus2pt minus2pt}
\def\bigskip{\vskip12pt plus4pt minus4pt}
\def\smallskip{\vskip3pt plus1pt minus1pt}
\centerline{\bf The Angle Resolved Photoemission Pseudogap and Anomalous
Background}
\centerline{\bf of Underdoped Bi$_2$Sr$_2$CaCu$_2$O$_{8+\delta}$
as Evidence for a Fermi Level Band Crossing}
\bigskip
\centerline{Jason K. Perry and Jamil Tahir-Kheli}
\centerline{\it First Principles Research, Inc.}
\centerline{\it 8391 Beverly Blvd., Suite \#171, Los Angeles, CA 90048}
\centerline{\it www.firstprinciples.com}
\bigskip
\centerline{Submitted to {\it Phys. Rev. Lett.}}
\bigskip
\noindent
{\bf Abstract.}  We show that the unusual observations of a pseudogap in the 
normal state of underdoped Bi$_2$Sr$_2$CaCu$_2$O$_{8+\delta}$ (BiSCO) using 
angle resolved photoemission spectroscopy (ARPES) is consistent with a new 
band structure for the cuprate superconductors in which the $x^2-y^2$ and 
$z^2$ bands are seen to cross at the Fermi level.  Limitations in the 
experimental method prevent the narrow 3D $z^2$ band from being fully 
resolved, leading instead to a broad background with ``stepfunction'' 
character.  As a consequence, the Fermi surface is mis-assigned and a 
pseudogap of approximately d-wave symmetry develops.
\vfill
\eject
A requirement of the BCS theory of superconductivity and its extensions is
the formation of a non-zero gap at the Fermi energy if and only if the material
is in the superconducting state $(T<T_c)$.
Thus the observation of an approximate $d_{x^2-y^2}$ gap in the
normal state $(T>T_c)$ of underdoped Bi$_2$Sr$_2$CaCu$_2$O$_{8+\delta}$
(BiSCO) using angle-resolved photoemission spectroscopy (ARPES)
is highly unexpected.$^1$  Present explanations for
this phenomenon speculate that above $T_c$ Cooper pairs are formed without 
long-range order of the wavefunction, producing a pseudogap
without superconductivity.  Below $T_c$, the wavefunction has long-range
phase order and superconductivity appears.  However a detailed
theoretical formulation of Cooper pairing without coherence is lacking.

In this paper, we argue
the ARPES pseudogap is {\it not} 
the result of Cooper pair formation above $T_c$, but
arises simply from the crossing of two bands at the Fermi level.
While conventional LDA band structure calculations do not predict the
existence of two such bands, we demonstrated in a series of articles
that correlation, which is well acknowledged to be missing in these
calculations, changes the band structure more radically than previously
assumed.$^2$  The new band structure is characterized by a narrow
3D $Cu\ d_{z^2}/O'\ p_z$ band ($z^2$) which crosses the broader, nearly
2D $Cu\ d_{x^2-y^2}/O\ p_{\sigma}$ band ($x^2-y^2$) at the Fermi level.
Our calculated band structure for optimally doped La$_{1.85}$Sr$_{0.15}$CuO$_4$
(LaSCO) is shown in Figure 1.  Significantly, a degeneracy of the two
bands is allowed by symmetry along the $(0,0)-(\pi,\pi)$ direction of 
the Brillouin zone.  Elsewhere in the Brillouin zone the bands repel.
This observation proves to be the essential ingredient in the Interband
Pairing Theory (IBP) of high temperature superconductivity.$^3$  
This theory postulates
the formation of a new type of Cooper pair (interband pair) comprised
of a $k\uparrow$ electron from one band and a $-k\downarrow$ electron
from another band in the vicinity of the $x^2-y^2$ and
$z^2$ symmetry allowed crossing.
As described in the above references, IBP can
explain a broad range of experimental data from simple band structure
arguments.  Such experimental data includes the observed
d-wave Josephson tunneling (and by extension the ARPES d-wave superconducting
gap), the temperature dependence of the Hall effect, the NMR,
the mid-IR absorption, and the incommensurate neutron scattering.$^3$

The origin of the pseudogap in our scenario is due to the difference
in the 3D dispersion of the $x^2-y^2$ and $z^2$ bands and the fact
that ARPES is a method
that maps a Fermi surface in 2D.  Electrons ejected from
$k$ states with predominantly $z^2$ character produce a broad linewidth
in ARPES and hence an unresolvable quasiparticle peak.
A mis-assignment of the Fermi surface results
and with it the pseudogap. This pseudogap is a direct measure of 
$x^2-y^2/z^2$ band repulsion, which has approximately d-wave symmetry.
From these considerations,
described in detail below, we conclude
the pseudogap is evidence for a Fermi level band crossing and is
unrelated to the d-wave superconducting gap.
We further show by calculation
that the anomalous background spectrum present in
all cuprate ARPES data is due to both primary and inelastically scattered
secondary electrons associated with the narrow $z^2$ band.

ARPES on optimally doped and overdoped BiSCO
yields a single holelike Fermi
surface closed around $(\pi,\pi)$.$^4$  At temperatures
below $T_c$,
the Fermi surface exhibits a d-wave gap with a node along the
$(0,0)-(\pi,\pi)$ direction.$^5$  This gap disappears isotropically as the
temperature is increased to $T_c$, as one would expect for a 
superconducting gap.  For underdoped BiSCO, the situation is
different as follows:

\itemitem{1.)}  The gap does not appear to be purely d-wave.$^6$

\itemitem{2.)}  The temperature dependence of the gap is anisotropic.$^7$

\itemitem{3.)}  The gap grows larger as the material is further
underdoped and $T_c$ is decreased.

\itemitem{4.)}  The gap persists through temperatures above $T_c$.$^1$

\noindent
Based on our calculated band structure for LaSCO, we conclude the following:

\itemitem{1.)}  The $x^2-y^2$
band is broadly dispersing in the CuO$_2$ planes ($x$ and $y$
directions) and narrowly dispersing normal to the
planes ($z$ direction).  It is an approximately 2D band.

\itemitem{2.)}  The $z^2$ band is narrowly dispersing in the $x$ and $y$
directions and moderately dispersing in the $z$ direction.  It must be
considered a 3D band.

\itemitem{3.)}  The two bands cross at or near the Fermi level.  The
crossing is allowed by symmetry along the $(0,0)-(\pi,\pi)$ diagonal but 
avoided elsewhere in the Brillouin zone.

In order to reconcile this proposed band structure with the Fermi surface
mapped by ARPES, one must consider the limitations in the ARPES method.$^8$
The experimentally measured ARPES linewidth $\Gamma_m$,
is a combination of the linewidth or lifetime of the photohole
$\Gamma_{hole}$
(hole state left behind by the excited electron) and the
linewidth of the photoelectron $\Gamma_{elec}$ (excited state
of electron after photon absorption). It is given approximately
by,

$$\Gamma_m = \Gamma_{hole} +
(v_{i,z}/v_{f,z})\Gamma_{elec},\eqno(1)$$

\noindent
where 
$v_{i,z}$ is the average Fermi velocity of the photohole in the $z$
direction, and $v_{f,z}$ is the average Fermi velocity of the photoelectron
in the $z$ direction.
While $\Gamma_{hole}\rightarrow0$ at the Fermi level, 
$\Gamma_{elec}$ is generally greater than or on the order of $1.0$ eV.

For the very 2D $x^2-y^2$ band, dispersion in the $z$ direction is
negligible compared to that in the $x$ and $y$ directions. Hence,
$v_{i,z}$ is expected to be small relative to $v_{f,z}$.  Thus
$\Gamma_m\approx\Gamma_{hole}$, leading to a resolvable quasiparticle
peak and a well defined Fermi surface crossing
for $k$ states with predominantly $x^2-y^2$ character.

In contrast, the $z^2$ band is narrowly dispersing in the
$x$ and $y$ directions but has larger dispersion in the
$z$ direction.  For this band, there is no reason to expect 
that $v_{i,z}$ is not
comparable to $v_{f,z}$.  Thus, the contribution from the
linewidth of the photoelectron cannot be neglected,
leading to a broad peak cutoff by the Fermi function.  
For $k$ states with predominantly $z^2$ character, there will be
no resolvable quasiparticle peak but instead a signal that looks like
a step function.

Given that only the $x^2-y^2$ band leads to a well resolvable peak
with ARPES, the band structure for LaSCO
produces a Fermi surface shown in Figure 2, in excellent
agreement with recent observations for this material.$^9$  The $z^2$
band contributes only to a broad background signal.  
Such a signal has been a signature
in ARPES on the cuprates, but has always been regarded with
confusion.$^{4-8}$  
Careful experiments with light polarization and photon energy dependence
may resolve if this background is due to $z^2$
character.  Such experiments have been done to confirm that the
major resolvable peak is in fact due to $x^2-y^2$ character.$^{8,10}$
The analysis is unfortunately complicated by secondary inelastic scattering
processes which may indeed dominate the background signal.  Such processes
have previously been considered and dismissed as the source of the background
because the conventional band structure could not produce a large
enough signal and the observed step function character.$^{8,11}$
However, in Figure 3, we show the signal due
to inelastic scattering from our calculated band structure for LaSCO leads
to a background which is significantly larger than that obtained
from the conventional $x^2-y^2$ band structure.
The inelastic
scattering is proportional to the integrated density of states.
Thus, the signal is dominated by the very narrow $z^2$ band with
a large density of states within 0.1 eV of the Fermi level leading to
the correct step function character near the Fermi level. 
The final ARPES lineshape is composed of primaries from both the
$x^2-y^2$ and $z^2$ bands as well as secondaries.

The observation of a pseudogap in the cuprates occurs when the true
Fermi surface is dominated by $z^2$ character.  The locus of $(k_x,k_y)$
points that comprise the Fermi surface is determined by measuring spectra
as one scans through the Brillouin zone from occupied $k$ states to
unoccupied $k$ states.  A quasiparticle peak appears as one approaches
the Fermi surface and disappears as one scans through the Fermi surface.
With our band structure, the dispersing quasiparticle peak will collapse
in $k$ space when we cross from states with predominantly $x^2-y^2$
character ($i.e.$ narrow linewidth) to states with predominantly $z^2$
character ($i.e.$ broad linewidth).  The extent (and possibly position)
of the collapse of the quasiparticle peak is also subject to matrix
element effects, which are different for $x^2-y^2$ and $z^2$.
Should this crossover occur for $k$ states
below the Fermi level, the Fermi surface will be mis-assigned.  For lower
temperatures, the leading edge will be below the Fermi level, leading
to the appearance of a pseudogap.  At
sufficiently high temperatures the $x^2-y^2$ hole linewidth will be large
enough that the leading edge of the spectra will be at the Fermi level,
closing the pseudogap anisotropically with temperature.$^{7}$

In order to explain the pseudogap in BiSCO, we need to
qualitatively understand its band structure.
While we have so far only calculated the band structure for
LaSCO, simple topological arguments can be used to
understand BiSCO.  The
principal difference between the two materials is that BiSCO
has two CuO$_2$ planes per unit cell instead of one.
This means there will be bonding and antibonding combinations of
both the $x^2-y^2$ and $z^2$ bands leading to a total of four
key bands instead of two.  Of these, the two $x^2-y^2$ bands
are nearly degenerate (there is little $z$ axis coupling
between them), but the two $z^2$ bands should be reasonably split in energy
such that only three bands (the two $x^2-y^2$ bands and the antibonding
$z^2$ band) are important at the Fermi level.  The fourth bonding $z^2$
band should be lower in energy.

In Figure 4, we present a schematic of the dispersion of the three key bands
along the symmetry lines $(0,0)-(\pi,\pi)$ and $(\pi,0)-(\pi,\pi)$.
The shaded region in the figure for $z^2$ antibonding shows
the spread in the dispersion
of this band as a function of $k_z$.  The two $x^2-y^2$
bands will vary little as a function of $k_z$ due to their
approximately 2D character.
Along $(0,0)-(\pi,\pi)$, the different reflection symmetries
of $x^2-y^2$ versus $z^2$ allow the bands to cross.  This crossing, which
is crucial to IBP, can persist at the Fermi level
over a range of dopings due to the $z$-axis
dispersion of the $z^2$ band.  From
$(\pi,0)-(\pi,\pi)$, only the $x^2-y^2$ bonding and $z^2$ antibonding bands
can cross and only if $k_z = {\pi/c}$ or $k_z = 0$. 
Elsewhere there is no symmetry to forbid mixing and the three
bands must repel.  This repulsion has approximately d-wave symmetry.

Figure 5 shows the approximate characters of the three bands in a 2D
Brillouin zone.
One can see that the top band
has $x^2-y^2$ antibonding character near $(\pi,\pi)$ that changes
over to $z^2$ antibonding character at $(0,0)$ and $(\pi,0)$. 
The middle band has $x^2-y^2$ bonding character at $(\pi,\pi)$
that changes over to $z^2$ antibonding for a small region and then
changes again to $x^2-y^2$ antibonding character at $(0,0)$ and
$(\pi,0)$.  The lowest band is $z^2$ antibonding at 
$(\pi,\pi)$ that becomes $x^2-y^2$ bonding at $(0,0)$ and $(\pi,0)$.
For all relevant dopings of BiSCO, the Fermi surfaces of all three
bands will be dominated by $z^2$ character.

Following our argument above, ARPES will mis-assign the Fermi surface
as the locus of $(k_x,k_y)$ points where character changes from antibonding
$z^2$ to antibonding $x^2-y^2$ in Band 2, as indicated by the dashed line.
Along the 
$(0,0)-(\pi,\pi)$ diagonal, there are always $x^2-y^2$ states at
the Fermi level since the $x^2-y^2$ and $z^2$ bands can be degenerate
here.  This leads to a zero gap ({\it i.e.} node) along the diagonal.
Scanning from $(\pi,0)-(\pi,\pi)$, the $x^2-y^2$ states lie
{\it below} the Fermi level due to band repulsion, producing a gap.
Thus, the approximately d-wave pseudogap in BiSCO and related
materials is due to a simple mis-assignment of the Fermi surface.

In regard to the relationship between the pseudogap and the superconducting
gap, as argued elsewhere,$^3$
the symmetry of interband pair to BCS pair scattering
produces a d-wave superconducting gap which forces conventional BCS
scattering to adopt this phase.  The possibility of interband pair
to interband pair scattering should produce an additional gap at the
nodes, but this would be extremely difficult to observe with ARPES due
to its strong $k_z$ dependence.  The confusion as to the pseudogap
arises because when
the material is underdoped, the pseudogap is larger in magnitude than
the superconducting gap, completely obscuring its presence.  
As doping is increased, the 
pseudogap is expected to decrease in magnitude as the Fermi level nears
the region where Band 2 switches to $x^2-y^2$ antibonding character.
Simultaneously, the superconducting gap
is expected to increase in magnitude.  At some point the superconduting gap
is expected to be greater in magnitude than the pseudogap, thus obscuring its
presence.  Eventually the pseudogap will disappear entirely as the Fermi
surface becomes increasingly $x^2-y^2$-like.  This behavior is consistent
with that observed for underdoped, optimally doped, and overdoped BiSCO.

It is important to note that since bands change character smoothly,
we have not defined exactly what $k$ point is the crossover from
$x^2-y^2$ to $z^2$.  The crossover momentum is dependent on
the sizes of the $x^2-y^2$ and $z^2$ ejection matrix elements and
these clearly are dependent upon the incident photon energy.  Thus,
the controversy over recent observations of a different Fermi surface
for BiSCO when the photon energy is $\approx 33$ eV$^{12}$ are
not in direct contradiction with the older $\approx 20-25$ eV results, but
instead demonstrate that measuring Fermi surfaces using ARPES
is not completely straightforward when the relevant bands
include orbitals with real 3D dispersion.

The authors wish to thank D.S. Dessau, Y.-D. Chuang, and A.D. Gromko for
many stimulating discussions.

\bigskip
\noindent
{\bf References}

\noindent
$^1$A.G. Loesser, {\it et al.}, Science {\bf 273}, 325 (1996); H. Ding,
{\it et al.}, Nature {\bf 382}, 51 (1996).

\noindent
$^2$J.K. Perry and J. Tahir-Kheli, Phys. Rev. B {\bf 58}, 12323 (1998); 
J.K. Perry, J. Phys. Chem., in press (cond-mat/9903088); 
J.K. Perry and J. Tahir-Kheli, Phys. Rev. Lett., submitted 
(cond-mat/9907332).  See also {\it www.firstprinciples.com}.

\noindent
$^3$J. Tahir-Kheli, Phys. Rev. B, {\bf 58}, 12307 (1998); 
J. Tahir-Kheli, J. Phys. Chem., in press (cond-mat/9903105);
J. Tahir-Kheli, Phys. Rev. Lett., to be submitted;
J. Tahir-Kheli and J.K. Perry, to be published.  See also
{\it www.firstprinciples.com}.

\noindent
$^4$H. Ding, {\it et al.}, Phys. Rev. Lett. {\bf 76}, 1533 (1996).

\noindent
$^5$Z.-X. Shen, {\it et al.}, Phys. Rev. Lett. {\bf 70}, 1553 (1993);
H. Ding, {\it et al.}, Phys. Rev. B {\bf 54}, 9678 (1996).

\noindent
$^6$J. Mesot, {\it et al.}, to be published (xxx.lanl.gov/cond-mat/9812377).

\noindent
$^7$M.R. Norman, {\it et al.}, Nature {\bf 392}, 157 (1998).

\noindent
$^8$N.V. Smith, P. Thiry, and Y. Petroff, Phys. Rev. B {\bf 47}, 15476 (1993);
Z.-X. Shen and D.S. Dessau, Phys. Rep. {\bf 253}, 2 (1995).

\noindent
$^9$A. Ino, {\it et al.}, J. Phys. Soc. Japan {\bf 68}, 1496 (1999).

\noindent
$^{10}$H.Ding, {\it et al.}, Phys. Rev. Lett. {\bf 76}, 1533 (1996).

\noindent
$^{11}$L.Z.Liu, J. Phys. Chem. Solids {\bf 52}, 1471 (1991).

\noindent
$^{12}$Y.-D. Chuang, {\it et al.}, Phys. Rev. Lett. {\bf 83}, 3717 (1999);
D.L. Feng, {\it et al.}, to be published (cond-mat/9908056);
H.M. Fretwell, {\it et al.}, to be published (cond-mat/9910221);
J. Mesot, {\it et al.}, to be published (cond-mat/9910430).

\vfill
\eject
\noindent
{\bf Figure Captions.}
\bigskip
\noindent
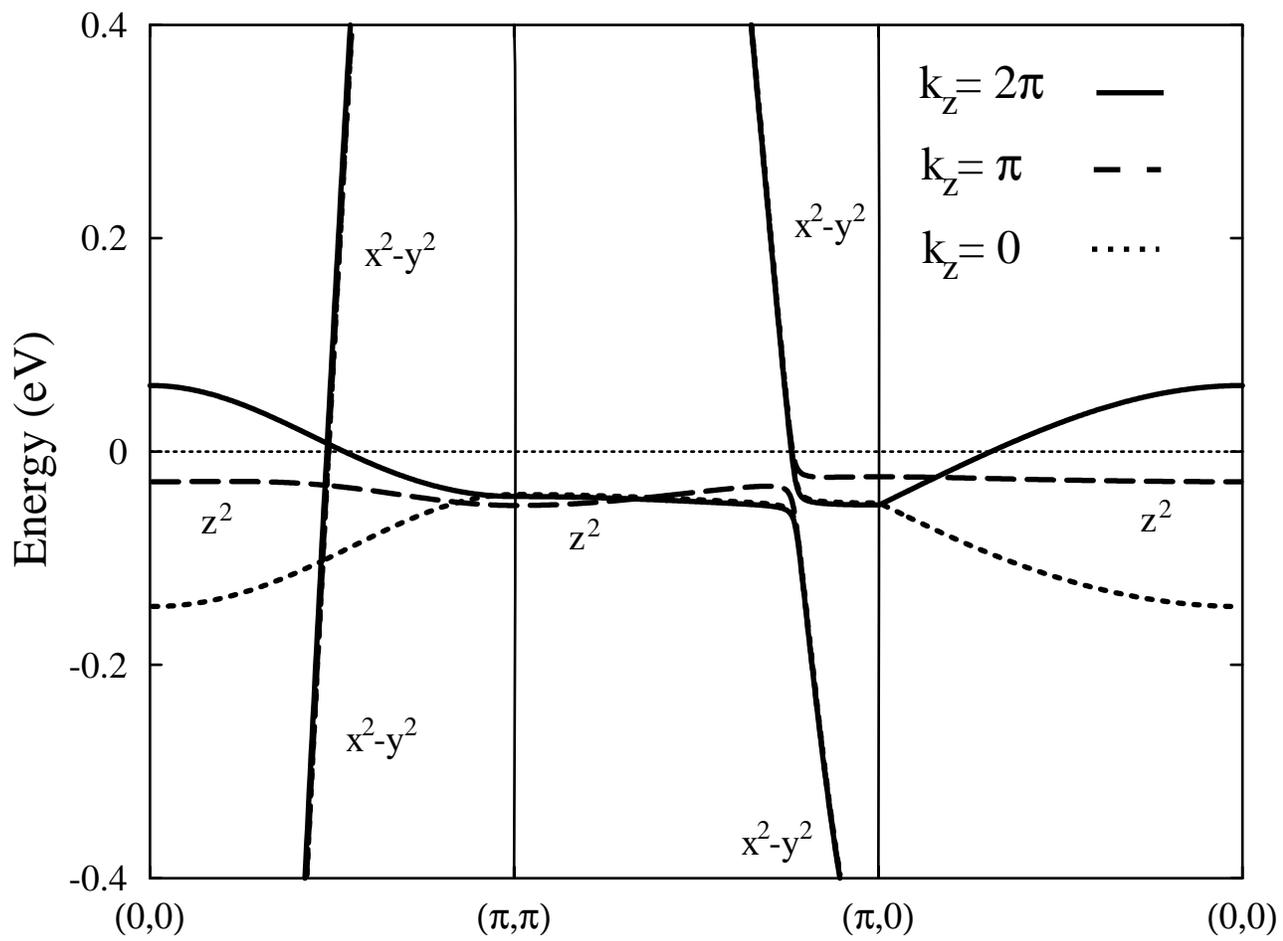
{\bf Figure 1.}  Calculated 3D band structure for optimally doped
LaSCO (see reference 2).  Band dispersion along
$(k_x,k_y)$ symmetry lines is given for $k_z = 0$, $\pi/c$, and
${2\pi}/c$.  Note the $x^2-y^2$ and $z^2$ bands cross along the
$(0,0)-(\pi,\pi)$ symmetry line but repel near $(\pi,0)$.

\noindent
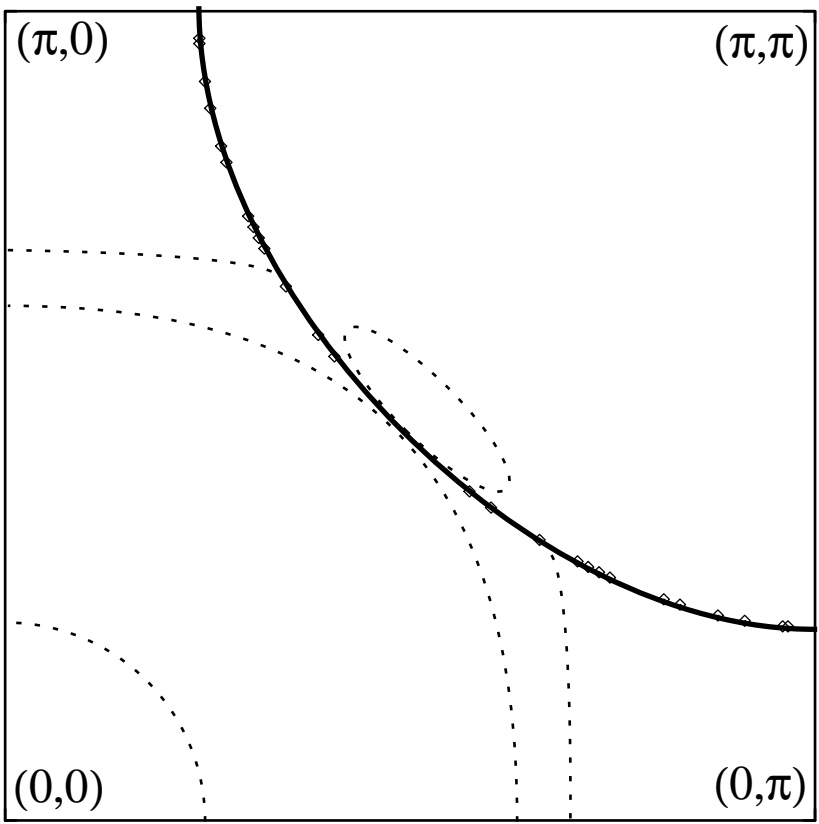
{\bf Figure 2.}  Calculated 2D Fermi surface (solid line) 
for optimally doped LaSCO that would be oberved by ARPES.
The true Fermi surface also contains 3D character, cross sections of which
are represented by the dotted lines.
This 3D component
contributes to a broad background signal with no resolvable Fermi surface in
the ARPES spectrum.

\noindent
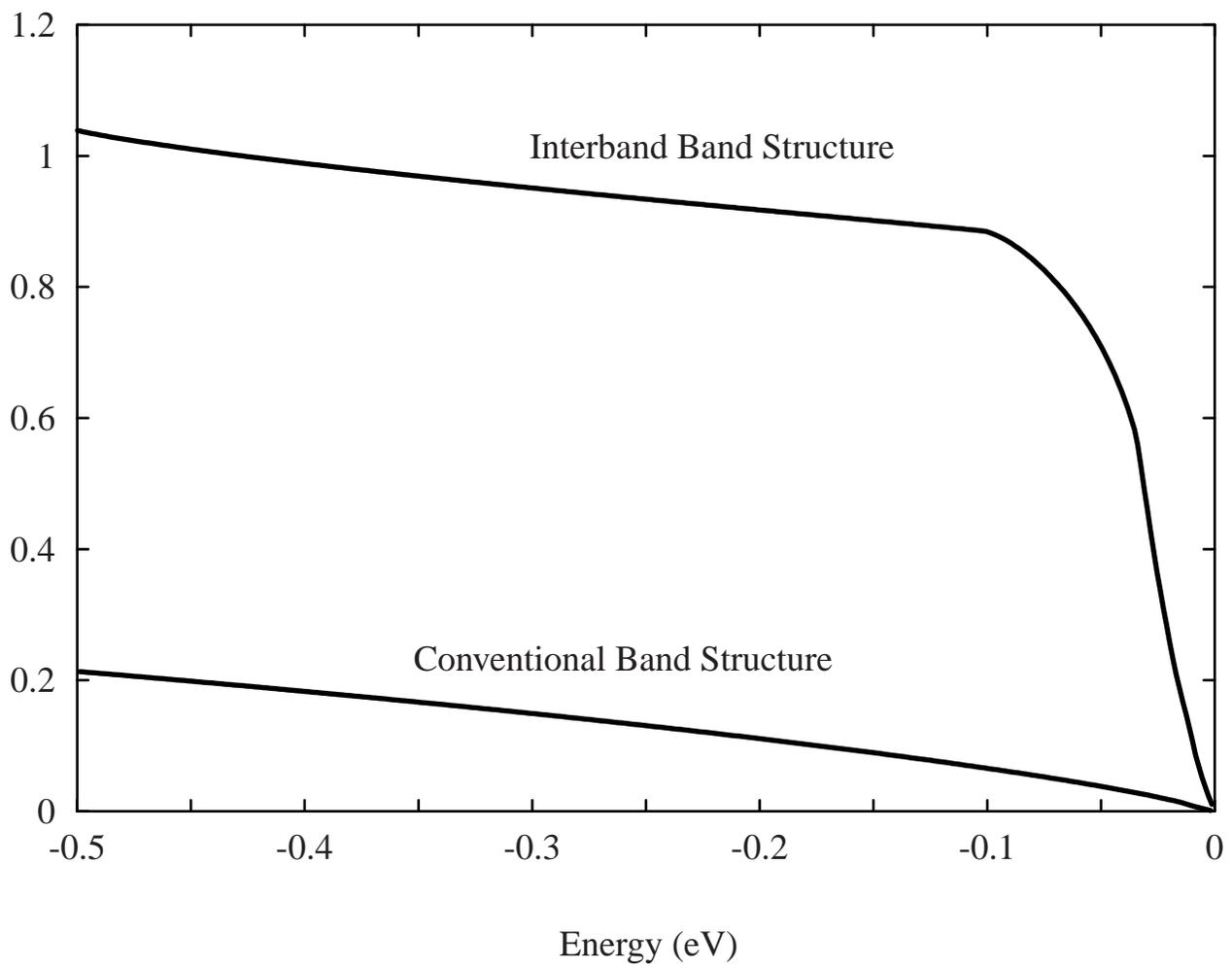
{\bf Figure 3.}  Integration over the occupied density of states for
our calculated band structure for LaSCO vs. a conventional ($x^2-y^2$
only) band structure.  The ARPES background signal due to inelastic scattering
is directly proportional to this curve.  The background signal predicted
from our band structure is in excellent agreement with that which is observed.

\noindent
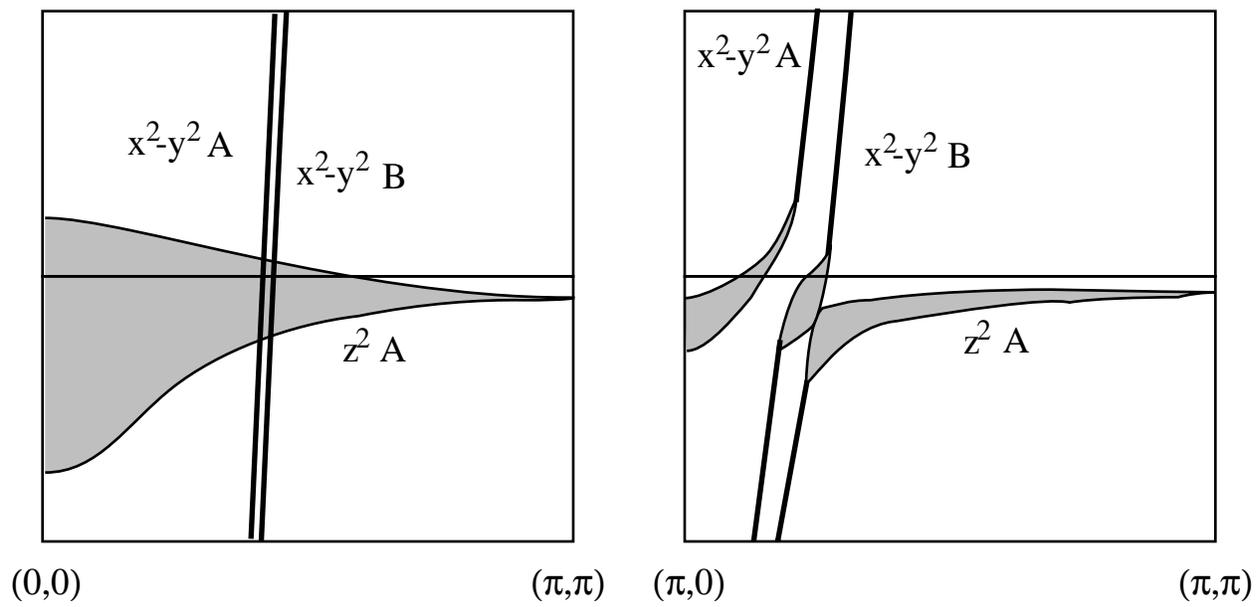
{\bf Figure 4.}  Schematic of the dispersion in BiSCO for
the $x^2-y^2$ bonding, $x^2-y^2$ antibonding,
and $z^2$ antibonding bands along the $(0,0)-(\pi,\pi)$
and $(\pi,0)-(\pi,\pi)$ symmetry lines.  The two $x^2-y^2$ bands cross
the $z^2$ band along the $(0,0)-(\pi,\pi)$ direction, but band repulsion
opens up an energy gap in the $x^2-y^2$ states along the $(\pi,0)-(\pi,\pi)$
direction.  The energy scale of the figure is $\approx$ 0.2-0.3 eV.

\noindent
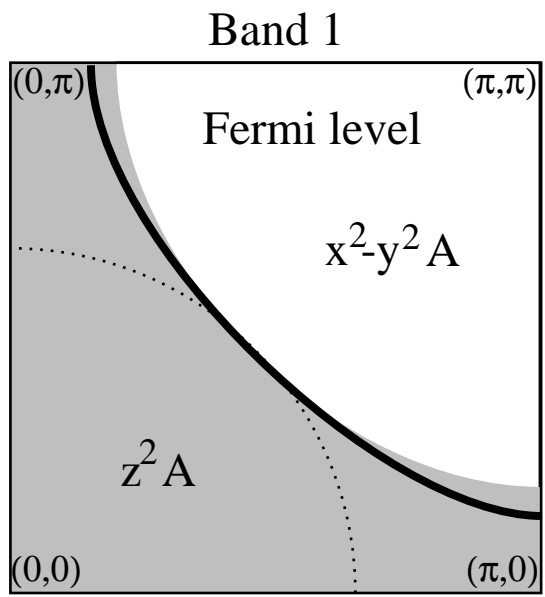
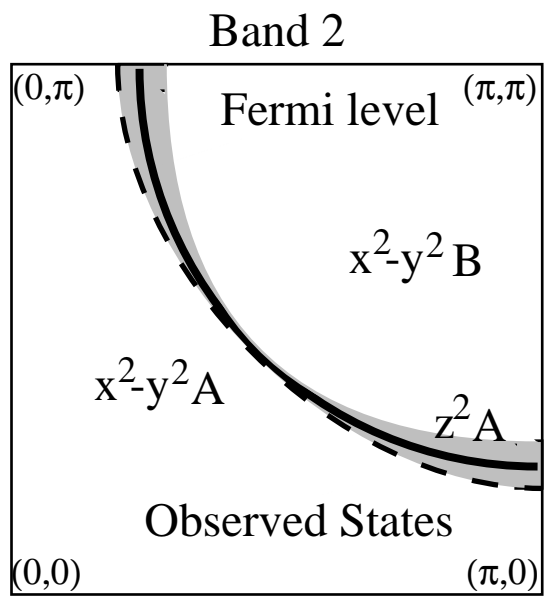
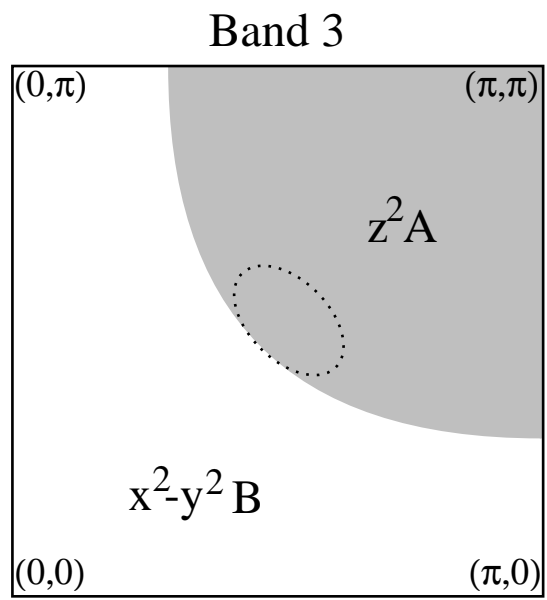
{\bf Figure 5.}  Schematic of the character of the three key bands in
BiSCO.  The true Fermi surface (which is approximately indicated by
the solid and dotted lines) has significant 3D character and cannot
be easily pinned down.  As a result,
ARPES mis-assigns the Fermi surface as indicated by the dashed line.
An approximate d-wave pseudogap is produced from this mis-assignment.
\end